\documentstyle[aps,twocolumn]{revtex}

\newcommand{\be}{\begin{equation}}
\newcommand{\ee}{\end{equation}}
\newcommand{\ea}{\end{eqnarray}}
\newcommand{\ba}{\begin{eqnarray}}

\begin{document}
\draft
\title{On the dimensional dependence of the electromagnetic duality groups}
\author{C.Wotzasek}
\address{Instituto de F\'\i sica\\
Universidade Federal do Rio de Janeiro\\
21945-970, Rio de Janeiro, Brazil}
\date{\today}
\maketitle
\begin{abstract}
We study the two-fold dimensional dependence of the electromagnetic duality groups.
We introduce the dual projection operation that systematically discloses the presence of an internal space of potentials where the group operation is defined.
A two-fold property of the kernel in the projection is shown to define the dimensional dependence of the duality groups.
The dual projection is then generalized to reveal another hidden two-dimensional structure.
The new unifying concept of the external duality space remove the dimensional dependence of the kernel, allowing the presence of both
$Z_2$ and SO(2) duality groups in all even dimensions.
This result, ultimately unifies the notion of selfduality to all D=2k+2 dimensions.
Finally, we show the presence of an unexpected duality between the internal and external spaces leading to a duality of the duality groups.
\end{abstract}

\section{Introduction}

This paper is devoted to study the dimensional dependence of the electromagnetic duality.
The problem with the space-time dimensionality is a crucial one that has obscured the understanding of duality with technicalities and misconceptions.
The distinction among the different even dimensions is manifest by the following double duality relation,

\begin{equation}
\label{10}
\mbox{}^{**}F = \cases{+F,&if $D=4k+2$\cr
	-F,&if $D=4k$\cr}
\end{equation}
where $*$ denotes the usual Hodge operation and $F$ is a $\frac D2$-form.  This leads to two, apparently independent, difficulties.  First, the concept of selfduality seems to be well defined only in twice odd dimensions, and not present in the twice even cases.  Second, (\ref{10}) apparently leads to separate consequences regarding the duality groups in these cases. 
The invariance of the actions in different
$D$-dimensions is preserved by the following groups,

\begin{equation}
\label{20}
{\cal G}_d= \cases{Z_2,&if $D=4k+2$\cr
	SO(2),&if $D=4k$\cr}
\end{equation}
which are called the ``duality groups". The $Z_2$ is a discrete
group with two elements.
The duality operation is characterized by a one-parameter SO(2) group of symmetry in D=4k dimensions, while for D=4k+2 dimensions it is manifest by a discrete $Z_2$ operation.  This leads to the prejudice that only the 4 dimensional Maxwell theory and its 4k extensions would possess duality as a symmetry, while for the 2 dimensional scalar theory and its 4k+2 extensions duality is not even definable. 
This two-fold dependence of duality with dimensionality has been tackled by algebraic methods, but its physical origin has not been questioned yet.

A solution for the first problem came with the recognition of a 2-dimensional internal structure hidden in the space of potentials\cite{DZ,DT,SS}.  Transformations in this internal duality space have extended the concept of selfduality to all even dimensions.  In the D=4 the explicitly self-dual Maxwell theory is known under the names of Schwarz and Sen\cite{SS}\cite{HG}, but this deep unifying concept has also been appreciated by others\cite{DZ,DT}.
Recently this author\cite{60} has developed a systematic method for obtaining and investigating different aspects of duality symmetric actions that embraces all dimensions.  It contains two basic points; a field redefinition and the soldering formalism\cite{MS}.
The redefinition of the fields in the first-order form of the action naturally discloses the 2-dimensional internal structure hidden into the theory.  It is crucial to note that this procedure produces two distinct classes of dual theories characterized by the opposite signatures of the (2x2) matrices in the internal space.  These actions correspond to selfdual and anti-selfdual representations of the original theory. An interesting duality between these duality symmetric actions has been reported recently\cite{DS}.
It is easy to visualize how the internal space effectively 
unifies the selfduality concept in the different $4k+2$
and $4k$ dimensions. The dual field is now defined to include the internal
index $(\alpha, \beta)$ in the fashion,

\begin{equation}
\label{30}
\widetilde F^\alpha = \cases{\epsilon^{\alpha\beta}\mbox{}^{*}F^\beta,&if $D=4k$\cr
	\sigma_1^{\alpha\beta}\mbox{}^{*}F^\beta,&if $D=4k+2$\cr}
\end{equation}
where $\sigma_k$ are the usual Pauli matrices and 
$\epsilon_{\alpha\beta}$ is the fully
antisymmetric $2\times 2$ matrix with $\epsilon_{12} =1$. Now, 
irrespective of the
dimensionality, the double dual operation yields,
\begin{equation}
\label{40}
\widetilde{\widetilde F} = F
\end{equation}
which generalizes (\ref{10}).  Although the notion of selfduality is now well define in all D=2k+2 dimensions, it is quite evident that it behaves differently in the twice even and twice odd cases. This can be seen equivalently from the duality groups (\ref{20}) or from the actions that originate them, which leads us directly to the second problem mentioned above.

Indeed the dichotomy (\ref{20}) seems to be of much deeper physical origin, since it attributes different group structures to distinct dimensions.  However, an important observation has been made in \cite{60,BW} that calls for the unification of the duality groups for all even dimensions.
The meaningful point to observe is that duality symmetry is not restricted to string and field theories, but is also present in the simple one dimensional oscillator.  It is precisely the dual symmetry hidden in this system that permeates to all field theoretical examples.
The two duality symmetric actions induced by the field redefinition behave, both classically and quantically, as 2-dimensional chiral oscillators of opposite signatures.  These systems display the usual SO(2) symmetry that characterizes the 4k dimensions.  However, we have also identified a $Z_2$ transformation, typical of 4k+2 dimensions, that swaps the two chiral oscillators.  It is at this stage that the soldering formalism becomes important. The idea of soldering, introduced by Stone, is the natural framework to fuse together different aspects of a symmetry and the fields that represent it.  By soldering these two systems together, we have found a new one that naturally incorporates both the discrete and the continuous groups into the same system: the 2-dimensional oscillator.
This result breaks with the usual lore represented by (\ref{20}).
As mentioned in \cite{60}, since the soldered 2-dimensional oscillator manifests both these symmetries its corresponding group is O(2).  The soldering formalism has led therefore to a new unifying structure.
The extension of this procedure to higher even dimensions has been presented in \cite{BW} with analogous results.  Incidentally, the manifestly duality symmetric actions found in \cite{BW} were also shown to be Lorentz invariant.  It is important to mention that although the soldering of
SO(2) with
$Z_2$ into the O(2) group, together with the master action that carries
its representation clearly makes it manifest the presence of both group structures, the understanding of the physical nature of this phenomenon,
once again, becomes clouded by the technical details behind the formalism.
In a recent report\cite{BC}, the double group structure mentioned above
was also revealed, not at the level of the master action but in the individual linearized lagrangean by exploiting the language of modes and the nature of the one-dimensional complex oscillator.

It is the main goal of this paper to study the dimensional dependence
of the duality groups. 
The misconceptions discussed above are shown to be related. The dimensional dependence of the duality groups (\ref{20}), reveals a residual manifestation of the initial problem (\ref{10}) after incorporating the internal space.
In this regard the emergence of two new concepts are of fundamental importance.
In fact our results unveil the existence of another two-dimensional structure. We call this second two-dimensional space as the {\it external duality space}, for reasons that will become clear as we progress.
Once the transformations in the external duality space are taken into account, this new unifying concept, together with the internal duality space, will clearly reveal the simultaneous presence of $Z_2$ and $SO(2)$ as the duality groups in all even dimensional cases, generalizing (\ref{20}).
The group structure for each D=2k+2 will depend now from which space, internal or external, the duality is being considered.
The duality group dependence with distinct dimensions will be shown to be given as,\vspace{.5cm}

\noindent (i) D=4k+2

\begin{equation}
\label{50}
{\cal G}_d= \cases{Z_2,&for $\mbox{internal duality}$\cr
	SO(2),&for $\mbox{external duality}$\cr}
\end{equation}

\noindent (i) D=4k

\begin{equation}
\label{60}
{\cal G}_d= \cases{Z_2,&for $\mbox{external duality}$\cr
	SO(2),&for $\mbox{internal duality}$\cr}
\end{equation}
It is an interesting result that the duality group dependence with the dimensionality coming from the external space is intertwined with respect to the dependence coming from the internal space. We can effectively identify the intertwining operation as {\it a duality of the duality groups},

\begin{eqnarray*}
\mbox{internal duality} &:& Z_2\Longleftrightarrow\mbox{external duality}:SO(2)\\
\mbox{internal duality} &:& SO(2)\Longleftrightarrow\mbox{external duality}:Z_2
\end{eqnarray*}

Although the internal duality space had been discovered without a systematic derivation, to disclose the presence of the hidden {\it external duality space}, such a recourse is essential.
The systematic derivation of the 2-dimensional internal space presented in \cite{60,BW} is here extended to disclose to external space as well.
In section II we present this technique, here called dual projection\footnote{The dual projection is the natural extension to higher dimensions of the 2D chiral projection, introduced earlier\cite{NC} to study the coupling of chiral bosons to gravity.}, that defines the variables of the internal space and features the presence of a kernel.
The residual group property (\ref{20}) is clearly shown to originate from a two-fold property of the kernels, that is dimensionally dependent.
Although the use of local kernels is more traditional, nonlocal kernels are more appropriate for our purposes, leading to nonlocal actions. They will be carefully reviewed.
In Section 3 we generalize the dual projection in order to include to external space.  We will see then how the presence of this new space allows for a generalization of the kernels structure, producing a unification of the group structure. However, the new freedom introduced by the kernels will produce a surprising result.

\section{Dual Projection and Internal Space}

The dual projection operation, that systematically discloses the internal duality space of any theory in D-dimensions, will also be important to unhide the presence of the 2-dim external duality space, to be discussed in the next section.  In this section this operation is quickly discussed.
The distinction of the duality groups is manifest, in the dual projection approach by the following construction.  The first-order action for a free field theory is, in general, given as

\be
\label{70}
{\cal L} = \Pi \cdot \dot\Phi - \frac 12 \Pi\cdot\Pi - \frac 12 
\partial\Phi\cdot\partial\Pi
\ee
with $\Pi$ and $\Phi$ being generic free fields in D dimensions and $\partial$ an appropriate differential operator.
For visual simplicity we omit all tensor and space-time indices describing the fields, unless a specific example is considered.
The parity of $\partial$ has a particularly interesting dependence with the dimensionality,

\begin{equation}
\label{80}
P\left(\partial\right) = \cases{+ 1,&if $D=4k$\cr
	- 1,&if $D=4k+2$\cr}
\end{equation}
where parity is defined as,

\be
\label{90}
\int \Phi \cdot\partial\Psi = P\left(\partial\right)\int \partial\Phi\cdot\Psi
\ee
Take for instance the specific cases of 2 and 4 dimensions where $\partial$ is defined as,

\begin{equation}
\label{100}
\partial = \cases{\partial_x,&if $D=2$\cr
	\epsilon_{kmn}\partial_n,&if $D=4$\cr}
\end{equation}
which we recognize as odd and even parity respectively.
The internal space is disclosed by a suitable field redefinition as
$\left(\Phi\;,\; \Pi\right)\rightarrow\left(A_+\; ,\; A_-\right)$ that is dimensionally dependent,

\begin{eqnarray}
\label{110}
\Phi & \rightarrow & A_+ + A_-\nonumber\\
\Pi &\rightarrow & \eta\left(\partial A_+ - \partial A_-\right)
\end{eqnarray}
with $\eta=\pm$ defining the signature of the duality symmetric action.  The effect of the dual projection (\ref{110}) into the first order action is manifest as,

\begin{equation}
\label{120}
{\cal L} = \eta\left(\dot A_\alpha\sigma_3^{\alpha\beta}\partial A_\beta
+ \dot A_\alpha\epsilon^{\alpha\beta}\partial A_\beta\right) - \partial A_\alpha\cdot\partial A_\alpha
\end{equation}
We can appreciate the impact of the dimensionality over the structure of the internal space, and the role of the operator's parity in determining the appropriate group for each dimension.  For twice odd dimension the dual projection produces the diagonalization of the first-order action into chiral actions, while for twice even dimensions, the result is either a self or an anti-selfdual action, depending on the sign of $\eta$,

\begin{equation}
\label{130}
{\cal L} = \cases{\eta\dot A_\alpha\sigma_3^{\alpha\beta}\partial A_\beta - \partial A_\alpha\cdot\partial A_\alpha
  ,&if $D=4k+2$\cr
	\eta\dot A_\alpha\epsilon^{\alpha\beta}\partial A_\beta - \partial A_\alpha\cdot\partial A_\alpha
,&if $D=4k$\cr}
\end{equation}
By inspection on the above actions one finds that while the first is duality symmetric under the discrete $Z_2$, the second is invariant under the continuous one-parameter group $SO(2)$.  This is in accord with general discussion based on algebraic methods\cite{DGHT}. In fact, using the choice of operators in (\ref{100}) we easily find that the D=2 case describes a right and a left Floreanini-Jackiw chiral actions\cite{FJ} if we identify $\Phi$ with a scalar field,

\begin{equation}
{\cal L}= \eta\dot A_\alpha\sigma_3^{\alpha\beta} A'_\beta -  A'_\alpha\cdot A'_\alpha
\end{equation}
The second case, on the other hand, describes either selfdual or anti-selfdual Schwarz-Sen actions, according to the signature of $\eta$,
\begin{equation}
{\cal L}=\eta\dot \varphi_k^\alpha\epsilon^{\alpha\beta}B_k^\beta - B_k^\alpha\cdot B_k^\alpha
\end{equation}
if we identify $\Phi$ with a vector field, $\Phi \rightarrow \varphi_k^\alpha$ and $\partial\Phi\rightarrow \epsilon_{kmn}\partial_m \varphi_n^\alpha = B_k^\alpha$, with $B_k^\alpha$ being the magnetic field.

The general discussion above discloses, in full generality, that the dependence of the duality group with dimension is controlled by the parity of $\partial$.
This is an important observation that will help us to unhide the presence of the second duality space, responsible for the extension of both duality groups to all even dimensions. It is essential in what follows, to examine the possibility of a nonlocal projection, since this language is more appropriated for the global nature of the problem. Therefore, consider the following dual projection,

\begin{eqnarray}
\label{140}
\Phi(x) & \rightarrow & \int K(x,y)\left[A_+(y) + A_-(y)\right]dy\nonumber\\
\Pi(x) &\rightarrow & \eta\left(A_+(x) -  A_-(x)\right)
\end{eqnarray}
where the symbols introduced are self-explanatory.  The kernel of the projection is chosen so as to satisfy

\begin{equation}
\label{150}
\partial \cdot K(x,y)=\delta\left(x-y\right)
\end{equation}
The redefinition of variables induced by this dual projection gives,
\begin{equation}
\label{160}
L =\eta\int dx dy \left\{\dot A^\alpha K\sigma_3^{\alpha\beta}A^\beta
+\dot A^\alpha K\epsilon^{\alpha\beta}A^\beta\right\}
 - A^\alpha A^\alpha 
\end{equation}
To satisfy (\ref{150}), the local and nonlocal kernel's parity must be the same, i.e., $P(K)=(-1)^{\frac D2}$.
Once again the parity of the kernels is the two-fold property responsible ultimately for the dimensional dependence of the groups displayed by Eq.(\ref{20}).

It interesting to identify here the nonlocal formulation for the FJ model if, for D=2 the kernel is chosen as,

\be
\label{170}
K(x-y)=\frac 12 \epsilon(x-y)
\ee
that clearly satisfy (\ref{150}).  On the other hand, if, in the D=4 case, we define the kernel so that,

\be
\label{180}
\partial_k K_{mn}(\vec x - \vec y)= \frac 12 \epsilon_{kmn}\delta(\vec x - \vec y)
\ee
then (\ref{160}) corresponds to a nonlocal formulation of the Schwarz-Sen model,

\begin{eqnarray}
\label{190}
S &=&\eta\int\int d^4x\,d^4y\left\{\dot A_m^\alpha(x) K_{km}(\vec x -\vec y)\epsilon^{\alpha\beta} A_n^\beta(y)\right\}\nonumber\\
&&\mbox{}-\int d^4x \;A_k^\alpha(x) A_k^\alpha(x)
\end{eqnarray}
The use of the mapping,

\be
\label{200}
A_k^\alpha\rightarrow \epsilon_{kmn}\partial_m \varphi_n^\alpha
\ee
makes contact with its local version, described by a vector potential $\varphi_k^\alpha$.

Certainly, here too what is to be regarded as important in defining the appropriate group of duality is the parity of the kernel, so that the duality group in D=4k+2 is again $Z_2$  while that of D=4k is $SO(2)$ accordingly to the parity of the kernel being Odd or Even.  Notice that these are the only choices that do not violate condition (\ref{150}).
Clearly the present {\it status in quo} does not allow for an unification of the duality groups for all even dimensions.  However, as shown by the soldering formalism, it is possible to solder together the two groups into a larger group.  We shown next how a deep unifying structure emerges in the form of a new two-dimensional duality space that solves this conflict.

\section{Dual Projection and External Space}

Let us now study the origin and the consequences of the external duality space, that is responsible for the unification of the duality groups and the phenomenon of duality of the duality groups.
To disclose such structure we introduce {\it ad hoc} a two-dimensional basis, $\left\{\hat e_a(k,x), \; a=1,2\right\}$, with (k,x) being conjugate variables and the orthonormalization condition given as,

\be
\label{210}
\int dx \;\hat e_a(k,x)\hat e_b(k',x)=\delta_{ab}\delta(k,k')
\ee
We choose the vectors in the basis to be eigenvectors of the Laplacian,
$\nabla^2=\partial \cdot \partial$,

\be
\label{220}
\nabla^2 \hat e_a(k,x)= \lambda^2(k)    \hat e_a(k,x)
\ee
The action of $\partial$ over the $\hat e_a(k,x)$ basis is

\be
\label{230}
\partial \cdot \hat e_a(k,x)=\lambda(k) M_{ab}\hat e_b(k,x)
\ee
that together with definition (\ref{220}) gives\footnote{Here we use the matricial notation where $\left(\widetilde M\right)_{ab} = M_{ba}$.}

\be
\label{240}
\widetilde M\cdot M=I
\ee
Using this basis to represent the elementary fields, 
\begin{eqnarray}
\label{250}
\Phi(x)&=&\int dk\; q_a(k)\;\hat e_a(k,x)\nonumber\\
\Pi(x) &=& \int dk\; p_a(k)\; \hat e_a(k,x)
\end{eqnarray}
with $q_a$ and $p_a$ being the expansion coefficients, leads us to a representation of the action as a two-dimensional oscillator,

\be
\label{260}
S=\int dk \left\{p_a\dot q_a - \frac 12 p_a p_a -\frac {\lambda^2}{2} q_a q_a\right\}
\ee
Let us introduce the dual projection

\begin{eqnarray}
\label{270}
q_a(k) &\rightarrow & \kappa_{ab}\left(A^{(+)}_b(k) +A^{(-)}_b(k) \right)\nonumber\\
p_a(k) &\rightarrow & \mbox{}\eta \lambda\;\left(A^{(+)}_a(k) - A^{(-)}_a(k)\right)
\end{eqnarray}
Here the indices of the oscillator's coordinates $q_a$, characterize the {\it external duality space}, justifying its name.  The $\alpha,\beta=\pm$ indices characterizes, correspondingly the {\it internal duality space}.  The parameter $\eta=\pm$ gives the signature that characterizes the actions as (anti)self dual and $\kappa_{km}$ is the cherished kernel, now defined in the external space whose parity is responsible for the emergence of a surprising new result.  The effect of the dual projection (\ref{270}) over (\ref{260}) is,

\begin{equation}
\label{280}
L_\eta = \int \lambda \left\{\eta A^\alpha_a \sigma^{\alpha\beta}_3 \kappa_{ab}\dot {A}^\beta_b + \eta A^\alpha_a \epsilon^{\alpha\beta}\kappa_{ab}\dot{A}^\beta_b 
- \lambda A_a^\alpha A_a^\alpha\right\}
\end{equation}
The impact of the kernel's parity, as the two-fold property responsible for the dichotomy of the duality group concept, once again becomes evident, leading to two distinct duality symmetric actions.  Using the dimensional dependence given in (\ref{80}), $P(\kappa)=(-1)^{\frac D2}$ gives,

\noindent (i) For D=4k

\begin{equation}
\label{290}
L_\eta^E =\int \lambda\left\{\eta A^\alpha_a \epsilon^{\alpha\beta}\kappa_{ab}^E\dot{A}^\beta_b
- \lambda A_a^\alpha A_a^\alpha\right\}
\end{equation}

\noindent (ii) For D=4k+2

\begin{equation}
\label{300}
L_\eta^O = \int\lambda\left\{\eta A^\alpha_a \sigma^{\alpha\beta}_3 \kappa_{ab}^O\dot {A}^\beta_b - \lambda A_a^\alpha A_a^\alpha\right\}
\end{equation}
just like in (\ref{130}).  The point of departure here is the presence of two duality spaces.  So the structure of the group will depend on which set of indices we are examining the duality.  For the internal space, D=4k+2 gives a $Z_2$ group while the D=4k case displays a SO(2).  For the external space, we find just the opposite result.  These findings are expressed in (\ref{50}) and (\ref{60}). 
Observe that the group structure of the internal and external spaces are clearly complementary.  It is important to notice that this result is not
dependent on our choice of basis, expressed in (\ref{210}).
Although this choice for the external space basis would be
justified only by the final results, we were guided here by our experience
in soldering the representations of the SO(2) and $Z_2$ groups by 2-dimensional chiral oscillators.  Notice, in this respect, that the dimensionality of the external space is however unique.  It is dictated by the fact that a field may be represented by a collection of 1-dimensional oscillators and consequently, as seen in \cite{60}, as 2-dim chiral oscillators of opposite signature.

Now comes a crucial observation.  Let us recall that these duality symmetric actions were found for arbitrary dimensions.  Let us also recall that in functional space the kernel's parity are restrict by the rule $P(K)=P(\partial)=(-1)^{\frac D2}$, irrespective of being the dual projection a local or nonlocal one.  However the kernel $\kappa$ defined in the external space does not need to satisfy this rule.  Therefore, unlike (\ref{120}) or (\ref{160}), the actions (\ref{290}) and (\ref{300}) are valid for all D=2k+2 dimensions.  This generalizes the groups in (\ref{50}) and (\ref{60}) to,

\begin{equation}
\label{310}
{\cal G}_d^E= \cases{Z_2,&for $\mbox{internal duality}$\cr
	SO(2),&for $\mbox{external duality}$\cr}
\end{equation}

\begin{equation}
\label{320}
{\cal G}_d^O= \cases{Z_2,&for $\mbox{external duality}$\cr
	SO(2),&for $\mbox{internal duality}$\cr}
\end{equation}
for all D=2k+2 dimensions.  Here ${\cal G}_d^E$ and ${\cal G}_d^O$ originates from $L^E$ and $L^O$ respectively.
This is a new result that shows that at all even dimensions both internal and external indices manifest both the $Z_2$ and the SO(2) groups,
in agreement with the findings of \cite{BW}.

In fact a new structure is illuminated leading to a new symmetry. 
Making the following choices of kernels, in terms of Pauli matrices

\begin{equation}
\label{330}
\kappa = \cases{\kappa^E=\sigma_3 & $$\cr
	\kappa^O=\epsilon & $$\cr}
\end{equation}
the even and odd actions above become

\begin{equation}
\label{340}
L_\eta^E =\int \lambda\left\{\eta A^\alpha_a \epsilon^{\alpha\beta}\sigma^3_{ab}  \dot{A}^\beta_b
- \lambda A_a^\alpha A_a^\alpha\right\}
\end{equation}

\begin{equation}
\label{350}
L_\eta^O = \int\lambda\left\{\eta A^\alpha_a \sigma^{\alpha\beta}_3 \epsilon_{ab}\dot {A}^\beta_b - \lambda A_a^\alpha A_a^\alpha\right\}
\end{equation}
It is interesting to observe that a new duality has been developed between the internal and the external spaces for this special choice of kernels.  Indeed observe that by making the switch

\be
\label{360}
\left\{\alpha, \beta\right\}\Longleftrightarrow\left\{a, b\right\}
\ee
then 

\be
\label{370}
L^O \Longleftrightarrow L^E
\ee
This is a new phenomenon where the odd and even duality actions switch roles as the two-dimensional spaces get exchanged.  We call it as a duality of the duality actions.  Certainly, the roles of the duality groups with respect to their dependence with internal and external variables also get switched.

\section{Conclusion}

The extension of the selfduality concept from D=4k+2 to all D=2k+2 dimensions was an important step to extend the duality symmetry from the Hamiltonian to Lagrangian level.
A fundamental concept in this realization was the identification of a naturally defined internal, two-dimensional structure in the space of potentials.
The origin of this structure is traceable to the symmetry of the equations of motion and the Bianchi identity, $d F=d \mbox{}^* F=0$.
Since both of these relations are closed, each one of them call for the presence of a local potential, that form the basis for the internal space.
However at the Lagrangian level, selfduality is manifest differently in distinct dimensions.
There is a clear two-fold dependence with dimension that boils down to different groups for twice even and twice odd cases.
From the simple exercise of soldering, however it becomes clear that the two group structures have a common representation.

The search for the unification led us to examine the origins of the different duality actions and groups.
Since these actions are derivable from the dual projection,
the origin of this difference is traceable to the parity property of the kernel
involved in the projection.
This gives a clear indication of another hidden two-dimensional structure called as external duality space.
The inclusion of the external duality space generalizes the kernel involved in the dual projection.
Different duality symmetric actions were disclosed, displaying either $Z_2$ or SO(2) symmetry, depending on the kernel's parity.
Unlike the functional case, the kernel's parity of the dual projection including the external variables was shown not to be dimensionally dependent.
Therefore both duality groups are present, irrespective of the dimension being twice even or odd.
Also worth of mention is the presence of an interesting duality between the internal and external duality space for some special kernels. This is a new result that deserves further investigation.

\end{document}